\documentclass[twoside,twocolumn,british,english,preprintnumbers,nofootinbib]{revtex4}
\usepackage{amsmath}
\usepackage{amssymb}
\usepackage{graphicx}
\usepackage{epsfig,bbm}
\usepackage{babel}
\usepackage{color}
\usepackage{hyperref}

\def\be{\begin{equation}}
\def\ee{\end{equation}}
\def\bea{\begin{eqnarray}}
\def\eea{\end{eqnarray}}

\begin{document}

\title{Hierarchical Axion Inflation}

\author{Ido Ben-Dayan}

\author{Francisco Gil Pedro}

\author{Alexander Westphal}

\affiliation{Deutsches Elektronen-Synchrotron DESY, Theory Group, D-22603 Hamburg, Germany}

\begin{abstract}
We propose a new field theory mechanism for generating an effective trans-Planckian decay constant from sub-Planckian ones. Using the minimal two axions and a hierarchy between two axion decay constants is sufficient for realizing inflation through non-perturbative effects only and with minimal  tuning. The inflationary motion is kept entirely within a sub-Planckian domain. 
 We outline possible strategies of embedding the model in a string theory setup.
\end{abstract}

\preprint{DESY-14-065}

\maketitle

\section{Introduction}
The recent breathtaking progress involving precision cosmological observations, in particular of the CMB \cite{Hinshaw:2012aka, Ade:2014xna}, provides rather strong evidence for a very early phase of slow-roll inflation driven by a scalar potential $V(\phi)$. A very common mechanism to protect the flatness of the scalar potential against radiative corrections is to invoke some symmetry, most straightforwardly a shift symmetry $\phi \rightarrow \phi+const.$. Radiative stability then requires that the potential itself constitutes the 'order parameter' of the breaking of the shift symmetry. Hence if the order parameter vanishes, the potential vanishes and the symmetry is exact. Then by turning on the order parameter, one can control the potential and its possible corrections, rendering the inflationary model technically natural. 

An elegant suggestion along these lines is "natural inflation" \cite{Freese:1990rb}, where the inflaton is a pseudo Nambu-Goldstone boson, and the potential is simply $V=\Lambda^4(1-\cos(\phi/f))$, where $f$ is the axion decay constant. Here, there is still a residual shift symmetry of $\phi \rightarrow \phi+2\pi f$. For successful inflation one needs $f \gtrsim M_{\rm P}$, and in light of the recent BICEP2 results \cite{Ade:2014xna}, the axion decay constant has actually to be as large as $f\sim 10 M_{\rm P}$ to be in accord with a tensor to scalar ratio of $r \sim \mathcal O (0.1)$. $M_{\rm P}=2.4 \times 10^{18} \text{\ GeV}$ is the reduced Planck mass. The predictions of natural inflation for such large decay constants are rather similar to those of a free massive scalar field, $V=\frac{1}{2}m^2\phi^2$. Such a large axion decay constant is challenging from the effective field theory point of view, and demands a UV completion to make sense of. 

Specifically, we are interested in embedding such a scenario in string theory, where generically $f\ll M_{\rm P}$ exacerbating the challenge. This is because in Planck units, for all known cases $f\propto1/\mathcal{V}^m$ with $m>0$, 
where $\cal V$ is the volume of the compactified dimensions and for a controlled analysis ${\cal V} \gg 1$, \cite{Baumann:2014nda}.
Scenarios providing some UV completions at various levels of rigour have been proposed in \cite{ArkaniHamed:2003wu, McAllister:2008hb, Kim:2004rp, Berg:2009tg}. 

With an eye on string theory, two mechanisms stick out \cite{Kim:2004rp, Berg:2009tg}. In the Kim-Nilles-Peloso mechanism (KNP)  \cite{Kim:2004rp}, an effective $f_{eff}\gg M_{\rm P}$ is achieved by aligning two axion decay constants that are originally sub-Planckian, $f_i\ll M_{\rm P}$. This requires a precise cancellation between parameters to the level of $0.1-1\%$. Recently \cite{Choi:2014rja,Higaki:2014pja} have suggested generalizations of the alignment mechanism to more than two axions, partially relieving the tuning required in the original model, while~\cite{Kallosh:2014vja} presented a very minimal embedding of single-axion and KNP 2-axion inflation into supergravity, see also \cite{Kallosh:2007cc}.

In Dante's Inferno (DI) \cite{Berg:2009tg}, a generalization of the axion monodromy was proposed. The main idea was considering two canonically normalized fields/axions with $V=W(r)+ \Lambda_2^4(1-\cos(r/f_r-\theta/f_{\theta}))$. $W(r)$ is some monomial in $r$, generalizing the axion monodromy construction. A hierarchy between the decay constants $f_r \ll f_{\theta} \ll M_{\rm P}$ and the energy scale $\Lambda_2^4 \gg W(r_{in})$ allows one to integrate out a heavy mode, mostly $r$, thus giving rise to effectively single field chaotic inflation dynamics $V=W_{eff}(\phi_{eff})$. Thus, the fundamental axion decay constants are sub-Planckian and the entire inflationary dynamics are contained in a small region of field space of the fundamental fields $r,\theta$ with diameter $d_r\simeq f_r/f_{\theta}\Delta \phi_{eff} \ll M_{\rm P}$. Thus one avoids the need for "functional fine tuning" one generally encounters in the case of "large field models", $\Delta \phi \gg M_{\rm P}$, while keeping all the predictions of such models intact.

In this note, we combine the best features of both KNP and DI -- namely, we generate a parametrically super-Planckian effective axion decay constant from fields with sub-Planckian periodicity using only non-perturbative effects, while replacing the tuned alignment of KNP with a simple hierarchy similar to that of DI. The main tool in DI is the integrating out of the mixed cosine term with $f_r\ll f_{\theta}$. In our case however, $W(r)$ is another axion term of the form $\Lambda _1^4 \left(1-\cos\left[\frac{r}{f_{r_1}}\right]\right)$, which by itself still has a residual shift symmetry.  Actually $\Lambda_1,\Lambda_2$ can have arbitrary relative magnitude. The virtue is that first, $f_r\ll f_{\theta}<M_{\rm P}$ suffices and there is no need for an additional hierarchy between the summands in the potential. The motion of the two axions is then kept entirely within a sub-Planckian domain. Second, as in DI and contrary to KNP, there is no fine-tuned alignment in the axion decay constants. Third, such an additional axionic term supplemented by no requirements on the $\Lambda$'s is expected to be much easier to embed in a full string derived scenario.

\section{The model}
Consider the two axion model with the potential given by
\be
V= \Lambda _1^4 \left(1-\cos\left[\frac{r}{f_{r_1}}\right]\right)+\Lambda _2^4\left(1-\cos\left[\frac{r}{f_{r_2}}+\frac{\theta }{f_{\theta _2}}\right]\right), 
\label{eq:Vrtheta}
\ee
where $\Lambda_1$ and $\Lambda_2$ are arbitrary and all decay constants are sub-Planckian.
The mass matrix for this system at the origin, takes the form
\be
M^2=\left(
\begin{array}{cc}
 \frac{\Lambda _1^4}{f_{r_1}^2}+\frac{\Lambda _2^4}{f_{r_2}^2} & \frac{\Lambda _2^4}{f_{r_2} f_{\theta _2}} \\
 \frac{\Lambda _2^4}{f_{r_2} f_{\theta _2}} & \frac{\Lambda _2^4}{f_{\theta _2}^2}
\end{array}
\right).
\ee

Comparing the trace and the determinant of the mass matrix we find the exact relation
\be
\begin{split}
\frac{\text{tr}{M^2}}{\sqrt{\det{M^2}}}&\equiv \frac{m_1}{m_2}+\frac{m_2}{m_1}\\
&=\frac{f_{\theta _2} \Lambda _1^2}{f_{r_1} \Lambda _2^2}+\frac{f_{r_1} \Lambda _2^2}{f_{\theta _2} \Lambda _1^2}+\frac{f_{r_1} f_{\theta _2} \Lambda _2^2}{f_{r_2}^2 \Lambda _1^2}
\end{split}
\ee
This implies that in the limit of small $f_{r_2}$ this quantity blows-up, implying the vanishing of the smallest eigenvalue of the mass matrix, $m_1$, and signalling the emergence of a flat direction.

In the limit of small $f_{r_2}\ll f_{r_1}, f_{\theta_2}$ the eigenvalues of $M^2$ are
\be
m_1^2=\frac{f_{r_2}^2 \Lambda _1^4}{f_{r_1}^2 f_{\theta _2}^2},\qquad m_2^2=\frac{\Lambda _1^4}{f_{r_1}^2}+\Lambda _2^4\left(\frac{1}{f_{r_2}^2}+\frac{1}{f_{\theta _2}^2}\right) 
\ee
and so we see that the mass spectrum is hierarchical, with  $m_1\ll m_2$.
We stress that this result only relies on the the hierarchy between the decay constants and does not require any further tuning of the remaining parameters, namely of the $\Lambda$'s, as long as\footnote{If $\Lambda_2/\Lambda_1$ is smaller than this threshold, the expression of the $m_1^2$ changes to $m_1^2=\Lambda_2^4 / f_{\theta_2}^2$, satisfying $\lim\limits_{\Lambda_2\to 0} \det M^2=0$.}
\be\label{hierarch}
\frac{\Lambda_2}{\Lambda_1}>\sqrt\frac{f_{r_2}}{f_{r_1}}\quad.
\ee
This simple lower limit
on the allowed ratio of the $\Lambda$'s has the beneficial effect of allowing for an easier embedding into string theory.

To render the mass hierarchy more evident it is useful to rotate the field space basis by the matrix
\be
S=\frac{1}{\sqrt{f_{r_2}^2+f_{\theta _2}^2}}\left(
\begin{array}{cc}
 -f_{r_2} & f_{\theta _2} \\
 f_{\theta _2} & f_{r_2}
\end{array}
\right),
\ee
such that 
\be
\left(
\begin{array}{c}
r\\ \theta\end{array}
\right)=\ S\ \left(
\begin{array}{c}
\phi_1\\ \phi_2 \end{array}
\right).
\ee
This implies the following approximate relation between the two field space basis:
\be\label{eq:fieldrelation}
 r=\phi _2-\frac{f_{r_2} \phi _1}{f_{\theta _2}},\qquad \theta=\phi _1+\frac{f_{r_2} \phi _2}{f_{\theta _2}}\quad.
\ee

We would like to emphasize here, that the whole analysis above will be exactly the same if we replace the $\Lambda_1$-term in the potential by a general positive function $W(r)$ with $\partial_r^2 W>0$ and the $\Lambda_2$-term by a general positive function $g(r/f_{r_2}+\theta/f_{\theta_2})$ with $\partial_i^2 g>0$ in an ${\cal O}(f_{r_2},f_{\theta_2})$ region of $(r,\theta)$-values around the origin. Arranging for the enlarged field range by covering the slope of $W(r)$ with many 'terraced trenches' then requires $g(r/f_{r_2}+\theta/f_{\theta_2})$ to be a periodic function with periodicities $(2\pi f_{r_2},2\pi f_{\theta_2})$.

The relation eq.~\eqref{eq:fieldrelation} allows us to write the axionic potential of Eq. (\ref{eq:Vrtheta}) as 
\bea
V&=& \Lambda _1^4 \left(1-\cos\left[\frac{f_{r_2} \phi _1}{f_{r_1} f_{\theta _2}}+...\right] \right)\nonumber\\
&&\quad+\;\Lambda _2^4\left(1-\cos\left[\frac{\phi _2}{f_{r_2}}+...\right]\right),
\eea
where $...$ denote subleading terms in the $f_{r_2}$ expansion. Upon integrating out the heavy field $\phi_2$ one is left with a single axion model with enhanced  decay constant given by
\be
f_{eff}\equiv\frac{f_{r_1} f_{\theta_2}}{f_{r_2}}.
\ee
The crucial point is that a sufficient hierarchy $f_{r_2}\ll f_{r_1},f_{\theta_2}$ can easily make $f_{eff}$ super-Planckian while keeping the fundamental $f_i<M_{\rm P}$. In this limit we get $V\simeq \frac12 m_1^2\phi_1^2$ leading to $n_s=1-2/N_e\sim 0.97$, $r=8/N_e\sim 0.1$.

In table \ref{tab:1} we give three examples of sub-Planckian $f_{r_1}, f_{r_2}\text{ and } f_{\theta_2} $ leading to $f_{eff}\sim 10\ M_{\rm P}$ and with several orders of magnitude separating the light and heavy fields' masses. The large hierarchy in the mass spectrum allows for an effective single field description while the large $f_{eff}$ leads to inflation in the quadratic regime.

\begin{table}[h]
\begin{center}
\begin{tabular}{c|c|c||c|c|c}
$f_{r_1}$ & $f_{\theta_2}$ & $f_{r_2}$ & $m_1^2$ & $m_2^2$&$f_{eff}$ \\
\hline
$0.1$&$0.1$&$0.001$&$0.01$&$10^6$&$10$\\
$0.1$&$0.01$&$0.0001$&$0.01$&$10^8$&$10$\\
$0.01$&$0.1$&$0.0001$&$0.01$&$10^8$&$10$\\
\end{tabular}
\end{center}
\label{tab:1}
\caption{Illustrative examples of viable models of natural inflation in the quadratic regime. Decay constants in units of $M_{\rm P}$, masses in units of $\Lambda^4$ for $\Lambda_1=\Lambda_2$.}
\end{table}%

\begin{figure}[t]
\begin{center}
	\includegraphics[width=0.5\textwidth]{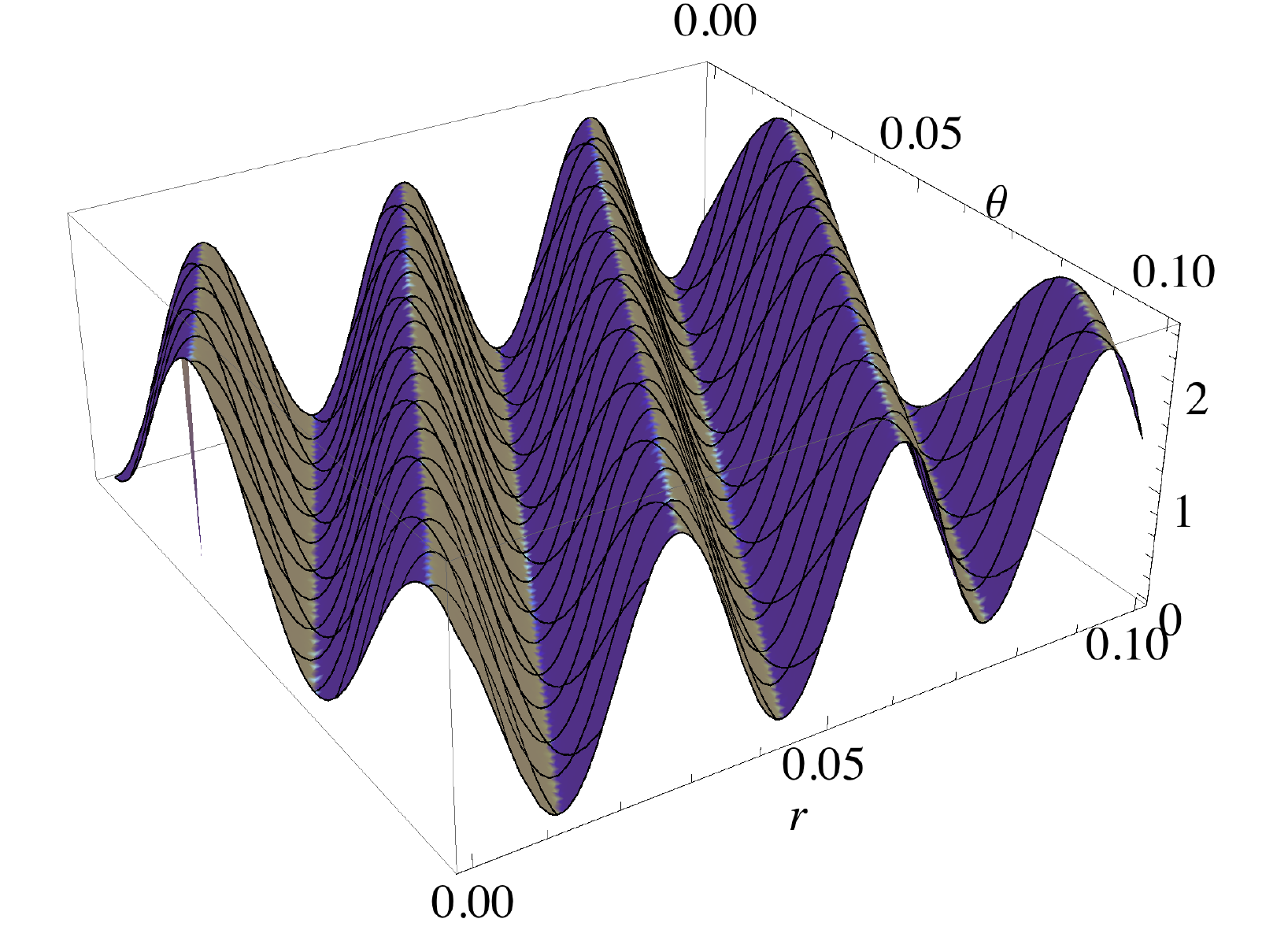}
\caption{The axionic potential in the $(r,\theta)$ coordinate system for the decay constants given by the second line of Table \ref{tab:1} in units of $\Lambda^4$ for $\Lambda_1=\Lambda=2$.}
\label{fig:Vrtheta}
\end{center}
\end{figure}

\begin{figure}[t]
\begin{center}
	\includegraphics[width=0.5\textwidth]{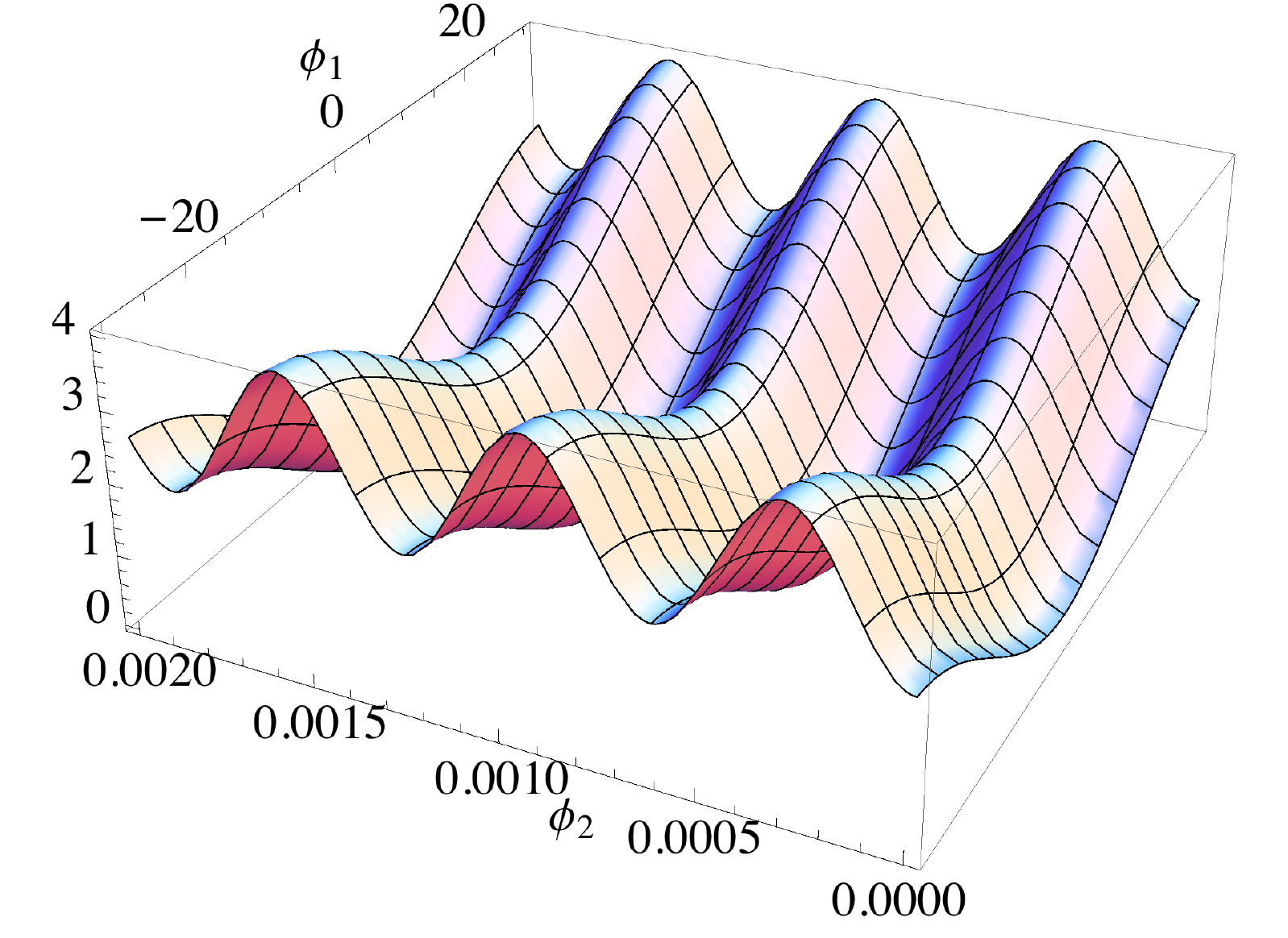}
\caption{The axionic potential of Fig. \ref{fig:Vrtheta} in the rotated coordinates $(\phi_1,\phi_2)$.  }
\label{fig:Vphi1phi2}
\end{center}
\end{figure}

In Fig. \ref{fig:Vrtheta} we plot the potential in the $(r,\theta)$ coordinate system for the decay constants of the second line of table \ref{tab:1}. It is clear that there is a light direction in V, which after field rotation can be identified with $\phi_1$, as shown in Fig. \ref{fig:Vphi1phi2}.

\section{Relation to previous work}

It is useful to point out the relations of our model here with previous attempts to generate large axion decay constants. We note, that our mechanism here combines the best features of both KNP~\cite{Kim:2004rp} and DI~\cite{Berg:2009tg}. Namely, we generate a parametrically super-Planckian effective axion decay constant from fields with sub-Planckian periodicity using only non-perturbative effects. Yet, we can do so without relying on the tuned alignment of KNP, employing instead just a simple hierarchy of decay constants similar to DI.

We can see the relation with the KNP setup by looking at its scalar potential
\bea
V_{KNP}&=& \Lambda _1^4 \left(1-\cos\left[\frac{m_1}{f_{r}}r+\frac{n_1}{f_{\theta}}\theta\right]\right)\nonumber\\
&&+\;\Lambda _2^4\left(1-\cos\left[\frac{m_2}{f_{r}}r+\frac{n_2}{f_{\theta}}\theta\right]\right) 
\label{eq:VKNP}
\eea
Upon rotating the field space into the eigenspace of the mass matrix, this potential can be rewritten as
\bea
V_{KNP}&=&\Lambda_1^4\,\left(1-\cos\left[\frac{\phi_h}{f_h}\right]\right)\nonumber\\
&&\quad+\;\Lambda_2^4\,\left(1-\cos\left[\frac{\phi_h}{\tilde f_h}+\frac{\phi_l}{f_l}\right]\right)
\eea
with effective decay constants $f_h$ , $\tilde f_h$, and $f_l$, with the latter given by
\be
f_l=\frac{\sqrt{m_2^2f_\theta^2+n_2^2f_r^2}}{|m_2n_1-n_2m_1|}\quad.
\ee
The KNP mechanism then consists of tuning an alignment $|m_1m_2-n_1m_2|\ll m_1^2,n_1^2$ to get $f_l\gg f_r, f_\theta$. We can see now, that our situation corresponds to a different regime of the same scalar potential which was not considered in~\cite{Kim:2004rp}. First, note that our case arises if we take $n_1\to 0$. Then
\be
f_l=\frac{\sqrt{m_2^2f_\theta^2+n_2^2f_r^2}}{n_2m_1}\quad.
\ee
Our situation corresponds now to 
\be
f_{r_1}\equiv\frac{f_r}{m_1}\;,\;f_{\theta_2}\equiv\frac{f_\theta}{n_2}\;\;\gg\;\; f_{r_2}\equiv\frac{f_r}{m_2}
\ee
which implies $m_2\gg m_1$ , $m_2\gg n_2 f_r/f_\theta$. In this limit we get
\be
f_l=\frac{m_2}{n_2m_1}f_\theta=\frac{f_{r_1}f_{\theta_2}}{f_{r_2}}\quad.
\ee
We see, that our situation arises from setting $n_1=0$ in $V_{KNP}$, upon which a simple hierarchy $f_{r_2}\ll f_{r_1}\,,\,f_{\theta_2}$ without any tuned alignment is enough to generate the large-field direction. The choice $n_1=0$ constitutes a simple discrete choice. This is different from a fine-tuning. To see this, we note that in the context of string theory constructions the linear combination of axion fields appearing in non-perturbative effects is governed by discrete topological quantities.

Hence, our mechanism takes the simplicity and control afforded by natural inflation incorporated in the KNP mechanism~\cite{Kim:2004rp}, and merges it with reduced amount of tuning from the Dante's Inferno scenario~\cite{Berg:2009tg} which the condition $f_{r_2}\ll f_{r_1}\,,\,f_{\theta_2}$ represents compared to the KNP mechanism. Thus, our model may well represent the realization of sub-Planckian periodicity  natural inflation with super-Planckian effective field range with: i) utilizing only non-perturbative effects, ii) the smallest number of axions, iii) and the least amount of tuning of the input parameters.

\section{Discussion and outlook}

Our discussion so far proceeded along the lines of 4D effective field theory. However, a model of natural inflation generating large-field directions with sub-Planckian decay constants and mild tuning requirements should ultimately be  embedded into string theory. Therefore, we now shortly discuss the ingredients which type IIB string theory with its well understood avenues for moduli stabilization, a pre-requisite for successful string inflation, supplies for axion inflation~\cite{Baumann:2014nda}. For the sake of concreteness, we will assume compactification of type IIB on a Calabi-Yau orientifold with 3-form flux, which supersymmetrically stabilizes the complex structure moduli and the axio-dilaton $\tau$ at a high mass scale, see ~\cite{Baumann:2014nda} and references therein. Axionic inflaton candidates arise from the RR 4-form and 2-form gauge potentials $C_4$ and $C_2$ respectively. $C_4$ provides $h^{1,1}_+$ axions as partners of the K\"ahler moduli $T_i = vol(\Sigma_i)+i\int_{\Sigma_i}C_4$. Here, $\Sigma_i$ are the $h^{1,1}_+$ 4-cycles surviving the orientifold projection. $C_2$ generates 2-form axions partnering with NSNS $B_2$ in $h^{1,1}_-$ orientifold-odd chiral fields $G_a=\int_{S^2_a}C_2-\tau \int_{S^2_a}B_2$. Here, $S^2_i$ denote the 2-cycles Poincare-dual to the $\Sigma_i$, while only the $a=1\ldots h^{1,1}_-$ orientifold-odd combinations $S^2_a$ lead to $G_a$ axion multiplets.


We now sketch two simple toy setups leading to an axion potential of the type of eq.~\eqref{eq:Vrtheta}. First, let us consider a simple model with three K\"abler moduli $T_i$ and their $C_4$-axion partners. We stabilize these moduli using non-perturbative effects from D7-branes or Euclidean D3-branes. Assuming for simplicity a 'swiss cheese' structure for the Calabi-Yau manifold, the effective action is governed by a K\"ahler and superpotential.
\bea
K &=& -2\,\ln\left[\left(T_L+\bar T_L\right)^{3/2}\hspace{-1.5ex}-\left(T_r+\bar T_r\right)^{3/2}\hspace{-1.5ex}-\left(T_\theta+\bar T_\theta\right)^{3/2}\right]\nonumber\\
&& \\
W\hspace{-1ex} &=& \hspace{-1ex}W_0\hspace*{-0.2ex}+\hspace*{-0.2ex}A_L e^{-\frac{2\pi}{N_L}T_L}\hspace*{-0.2ex}+\hspace*{-0.2ex}A_1 e^{-\frac{2\pi}{N_{r_1}}T_r}\hspace*{-0.2ex}+\hspace*{-0.2ex}A_2 e^{-2\pi \big(\frac{T_r}{N_{r_2}}+\frac{T_\theta}{N_{\theta_2}}\big)} \nonumber
\eea
Here $W_0$ is the vev of the flux superpotential after stabilization of the complex structure moduli and the axio-dilaton.
The F-term scalar potential provides KKLT-type minima for the moduli, while generating an axion potential for $r={\rm Im}\, T_r$ and $\theta={\rm Im}\, T_\theta$ of the kind of eq.~\eqref{eq:Vrtheta} after uplifting the KKLT minimum to approximately zero vacuum energy. A moderate hierarchy $N_L > N_r>1$ provides for stabilizing at moderately large volume ${\rm Re}\, T_L > {\rm Re}\, T_r>{\rm Re}\, T_\theta\sim {\cal O}(1)$. Alternatively, we may use a combination of $\alpha'$- and string-loop corrections to stabilize all or part of the real parts of the K\"ahler moduli~\cite{Baumann:2014nda}, while the perturbative nature of these corrections guarantees preservation of the $C_4$ shift symmetry.

A complete discussion of the tuning of the parameters and stability requires the inclusion of moduli stabilization and the canonical normalization effects originating from a non trivial metric in field space. For the moment we focus only on the level of potential tuning that is achievable only at the level of W, postponing a full discussion for \cite{Ben-Dayan:2014}. Choosing microscopic parameter values $N_L=30$ (corresponding to an $E_8$ D7-brane stack), $N_{r_1}=N_{\theta_2}=10$, $N_{r_2}=2$, $W_0\sim 0.01$, $A_L\sim1$, $A_1\sim0.01$, and $A_2\sim 10$, we see that the KKLT mechanism stabilizes the moduli at ${\rm Re}\, T_L\sim 25$~, ${\rm Re}\, T_r \sim 10$ , ${\rm Re}\, T_\theta \sim 2$. This leads via the exponential terms in $W$ to $\Lambda_1^4\propto W_0 A_1 \exp(-2\pi/N_{r_1} {\rm Re}\,T_r)\sim 10^{-5}$ and $\Lambda_2^4\propto W_0 A_2 \exp(-2\pi {\rm Re}\,(T_r/N_{r_2}+T_\theta/N_{\theta_2}))\sim 10^{-6}$, while providing for right hierarchy of the effective decay constants $f_{r_2}/f_{r_1}= N_{r_2}/N_{r_1} = 5$ and also  $f_{\theta_2}/f_{r_2}=N_{\theta_2}/N_{r_2}=5 $. Since ${\rm Re}\, T_L\sim 25$ implies a Calabi-Yau volume of ${\cal V}\sim 10^2\ldots 10^3$, the F-term scalar potential along the inflaton direction has a scale $\sim e^K\Lambda_2^4\sim 10^{-10}$ which is in the right range to satisfy COBE normalization of the inflationary curvature perturbations. The actual field range of the inflaton-axion potential will depend on the canonical normalization of the $\theta$ axion as $f_{eff}= 50 f^{kin}_{\theta} M_{\rm P}$ and so provided  $f^{kin}_{\theta}$ is not exceedingly small,  $f^{kin}_{\theta} \ge 1/50$, trans-Planckian field ranges are attainable.


An alternative setting can arise from utilizing the 2-form RR-axions.  A simple setup involving LVS volume stabilization, utilizing two D5-brane stacks and an Euclidean D3-brane (ED3)  looks like
\be
\begin{split}
K &= -2\,\ln\left[\left(T_L+\bar T_L\right)^{3/2} \right.\\ & \left.  -\left(T_s+\bar T_s+\sum_{a=r,\theta}c_a (G_a+\bar G_a)^2\right)^{3/2}-\hat\xi\right]
\end{split}\nonumber
\ee
\vspace{-5ex}
\bea
&&\\
&&\nonumber\\
W&= &W_0+A_s e^{-2\pi T_s}+A_1 e^{-\frac{2\pi}{N_{r_1}}G_r}+A_2 e^{-2\pi \big(\frac{G_r}{N_{r_2}}+\frac{G_\theta}{N_{\theta_2}}\big)}\nonumber
\eea
The first ED3 in $T_s$ fixes $T_s$ and the overall volume via $T_L$ by its interplay with $\alpha'^3$-correction parametrized by $\hat\xi\sim \chi(CY_3)$. Then the D5-brane stacks provide a scalar potential similar to eq.~\eqref{eq:Vrtheta} for the axions given by ${\rm Im}\,G_a=\int_{S^2_a}C_2$, while KKLT fixing the $B_2$-components of the $G_a$. We note that non-perturbative contributions to the K\"ahler potential arising from ED1 instantons can also generate the scalar potential for the $G_a$-axions~\cite{McAllister:2008hb}. In all cases, a crucial topological requirement for a string construction to reproduce an axion potential with a structure like eq.~\eqref{eq:Vrtheta} seems to be tied with the  engineering of a partially ample divisor depending on a linear combination of $T_r$ and $T_\theta$.

Summarizing, we have shown that a combination of two cosine potentials for two axions with a simple hierarchy of sub-Planckian decay constants can provide for successful large-field natural inflation with minimal tuning. This setup has the right bottom-up properties for potentially successful embedding into string theory construction, for which we discussed several possible avenues and which will be discussed in detail in \cite{Ben-Dayan:2014}.

\section*{Acknowledgments}
After completing this work we became aware of \cite{Tye:2014tja} discussing a similar approach to super-Planckian decay constants. However, the core model of this work has been announced in~\cite{Westphal:2014}. \\
 This work was supported by the Impuls und Vernetzungsfond of the Helmholtz Association of German Research Centres under grant HZ-NG-603. The work of I.B.-D. is supported by the German Science Foundation (DFG) within the Collaborative Research Center (CRC) 676 
``Particles, Strings,
 and the Early Universe''.

\end{document}